\documentclass[a4paper,12pt,oneside]{report} 

\usepackage{EMU_Thesis}
\usepackage[a4paper,left=4cm,right=2.5cm,top=2.5cm,bottom=2.5cm,footskip=.8cm]{geometry}
\pgfplotsset{width=10cm,compat=1.9}
\usepgfplotslibrary{external}

\title{Investigating the Generalized Uncertainty Principle Effects on
Hawking Radiation in Rotating Linear Dilaton Black Holes}
\author{Erdem Sucu}
\degree{Master of Science}
\inof{in}
\program{Physics}
\defenseyear{2024}
\defensemonth{January}
\Dept{Physics}
\InstituteDirector{Prof. Dr. Ali Hakan Ulusoy}
\DeptChair{Prof. Dr. İzzet Sakallı}
\supervisor{Prof. Dr. İzzet Sakallı}
\examiner{Assoc. Prof. Dr.}{Suat}{Dengiz}
\examiner{Asst. Prof. Dr.}{Huriye}{Gürsel Mangut}
\examiner{Prof. Dr.}{İzzet}{Sakallı}

\begin{document}

% page numbering for the preliminary sections
\pagenumbering{roman}

% Justify text without hyphenation 
% (You may disable this option but doing so may create problems in justifying paragraphs)
\tolerance=1
\emergencystretch=\maxdimen
\hyphenpenalty=10000
\hbadness=10000

%##################################################################
% Preliminary section
%##################################################################
%================== Title and Signature Pages =====================
\makethesistitle 
\makeapprovalpage

%================== ABSTRACT ======================================
\begin{abstract}
The impact of the Generalized Uncertainty Principle (GUP) on Hawking particle emission in a rotating linear dilaton black hole (RLDBH) spacetime is examined in this thesis. The concerned study presents a thermal emission model for black holes (BHs) that incorporates the influence of gravitational lens particles through GUP during the quantum tunneling process. The findings suggest that with GUP support, the temperature of Hawking radiation decreases as GUP parameters increase and rises with an increasing BH mass. The thesis also delves into the repercussions of these discoveries on the information loss paradox and adjusted entropy, while also exploring the potential utilization of astrophysical data to confirm GUP effects. In conclusion, our work underscores the significant role of GUP in the thermal emission of non-asymptotically flat (NAF), stationary BHs and its capacity to shed light on the intricate relationship between astrophysics and quantum gravity.

\end{abstract}

%================== ÖZ ============================================
\begin{ozet}
Bu tezde, genelleştirilmiş belirsizlik ilkesi'nin (GUP), dönen doğrusal dilaton kara delik uzayzamanında Hawking parçacık emisyonu üzerindeki etkisi incelenmektedir. Bu çalışma, kütleçekim mercek parçacıklarının GUP aracılığıyla kuantum tünelleme sürecindeki etkisini içeren kara delikler için bir termal emisyon modeli sunulmaktadır. Elde edilen bulgular, GUP desteği ile Hawking radyasyonunun sıcaklığının GUP parametreleri arttıkça azaldığını ve artan siyah delik kütleleri ile arttığını önermektedir. Tez ayrıca bu keşiflerin bilgi kaybı paradoksu ve ayarlanmış entropi üzerindeki etkilerini incelemekte ve aynı zamanda GUP etkilerini doğrulamak için astrofizik verilerinin potansiyel kullanımını araştırmaktadır. Sonuç olarak, çalışmamız, GUP'un asimptotik olarak düz olmayan, durağan kara deliklerin termal emisyonunda oynadığı önemli rolü ve astrofizik ile kuantum kütleçekimi arasındaki karmaşık ilişkiyi aydınlatma kapasitesini vurgulamaktadır.
 
\noindent \textbf{Anahtar Kelimeler}: Hawking Radyasyonu, GUP, Doğrusal Dilaton, Kuantum Tünelleme, Kara Delikler
\end{ozet}

%================== ACKNOWLEDGEMENTS (optional) ===================
\begin{acknowledgements} 
Having reached the final steps of completing my master's studies, there exist numerous individuals whom I wish to thank. Firstly, I am sincerely grateful to my family for their unwavering encouragement and support throughout my academic progress. Their love, compassion, and faith have been the driving main force behind my achievements.

I want to extend my deepest gratitude to my thesis supervisor, Prof. Dr. İzzet Sakallı, who played a crucial role in overseeing my research. His unwavering dedication to mentoring and academic excellence was instrumental in shaping this study. His insightful feedback and invaluable guidance helped me navigate the intricacies of this research, for which I am profoundly thankful.

Furthermore, I would like to thank Prof. Dr. Mustafa Halilsoy for sharing his experiences with me and inspiring discussions.

Lastly, I would like to express my heartfelt gratitude to the numerous researchers, scholars, and authors whose dedicated work served as the cornerstone of this study. Their significant contributions to the field have been immeasurable. I am profoundly appreciative of all the support and guidance I have been fortunate to receive, and I eagerly anticipate the opportunities that await in the future.

\end{acknowledgements}

%================== PREFACE (optional) ============================
%\begin{preface} 
% Preface is optional.
%\end{preface}

%================== TABLE OF CONTENTS =============================
\tableofcontents

%================== LIST OF FIGURES (if available) ================
\listoffigures

%================== ABBREVIATIONS (if available) ==================
% For the list of symbols and abbreviations you have two options:
% 1. You can use nomencl package for the list of symbols and abbreviations. Follow the given example below:
% Note: use [aa] prefix before English symbols to force English letter symbols sorted at the beginning, and use [ab] prefix to sort Greek letter symbols afterward. For abbreviations you don't need to use any prefix.

%\nomenclature[aa]{$h$}{Planck constant}
%\nomenclature[aa]{$c$}{Speed of light in a vacuum inertial frame}
%\nomenclature[ab]{$\phi$}{Golden ratio}
%\nomenclature[ab]{$\mu$}{Micro sign}
%\nomenclature{sps}{Samples Per Second}
%\nomenclature{IMU}{Inertial Measurement Unit}
%\nomenclature{SoC}{System on Chip}
%\nomenclature{DMP}{Digital Motion Processor}
%\nomenclature{DSP}{Digital Signal Processor}
%\nomenclature{QFN}{Quad Flat No-leads}
%\nomenclature{LED}{Light-Emitting Diode}
%\nomenclature{GPIO}{General Purpose Input/Output}

%\printnomenclature

% 2. You can also use the below definitions for symbols and abbreviations instead of nomancl package (explained above). This way you can have separate list of symbols and list of abbreviations (optional).
% It is recommended to sort the list by an external software. The internal sorting function may not be accurate depending on the content.

% Create symbols
\DTLnewdb{symbols}
\addsymbol{$l_{P}$}{Planck Length}
\addsymbol{$\hbar$}{Planck Constant}
\addsymbol[b]{S and $\mathcal{I}$}{ Tunnelling Action}
\addsymbol[b]{$\kappa$}{Surface Gravity}
\addsymbol{$\Omega$}{Angular Velocity}
% Sort the list if needed
\DTLsort*{Acronym}{symbols}

% Create abbreviations
\DTLnewdb{abbreviation}
% Define the ABBREVIATIONS database as follows
\addabbreviation{GUP}{ Generalized Uncertainty Principle}
\addabbreviation{QGC}{Quantum Gravity Corrected}
\addabbreviation{RLDBH}{Rotating Linear Dilaton Black Hole}
\addabbreviation{NAF}{Non-asymptotically Flat }
\addabbreviation{HUP}{Heisenberg Uncertainty Principle }
\addabbreviation{EMDA}{Einstein-Maxwell-Dilaton-Axion }
\addabbreviation{WKB}{Wentzel-Kramers-Brillouin}

% Sort the list if needed
\DTLsort*{Acronym}{abbreviation}

% Use one of the page titles below depending on the content of your list:

\begin{symbolsandabbreviations}

  % Display the contents of the database

\end{symbolsandabbreviations}

%\begin{symbols}
  % Display the contents of the database
%\end{symbols}

%\begin{abbreviations}
  % Display the contents of the database
%\end{abbreviations}

%##################################################################
% End o"

\chapter{INTRODUCTION}\label{ch:introduction}
Black holes (BHs) \cite{chandrasekhar1998mathematical} have captured the imagination of physicists due to their extraordinary properties and the potential they hold for shedding light on fundamental questions in physics, particularly those related to the nature of spacetime and the behavior of matter under extreme conditions. Hawking radiation explaining the escape of particles from BHS through quantum phenomena remains at the centre of the study of BH thermodynamics \cite{bardeen1973four,hawking1975particle,sakalli2017hawking,sakalli2014gravitational,gecim2014hawking,Gecim:2019pft}. The precise mechanism behind this radiation is still under investigation, leaving a variety of open questions in the scientific literature.

\pagenumbering{arabic}

The generalised uncertainty principle (GUP) \cite{scardigli1999generalized, das2008universality, amelino2006black, sakalli2022physical, gecim2018gup, sakalli2016gup}, which arises from the junction of quantum physics and gravity, modifies Heisenberg's uncertainty principle (HUP). GUP places a fundamental limitation on our ability to accurately measure certain pairs of properties such as position and momentum. In the last few years, there has been growing interest in exploring the influence of GUP on the behavior of BHs, as evidenced by the references \cite{heidari2023quantum, pourhassan2020pv, carr2022generalized, ovgun2016entangled, javed2019hawking, baruah2023quasinormal}. GUP effects alter the emission of Hawking radiation and impact the dynamics of BHs, as discussed in detail in \cite{sakalli2022topical} and its related sources. Moreover, it has been demonstrated that GUP influences the entropy of BHs and may be able to resolve the information loss dilemma, a persistent problem in theoretical physics \cite{chen2015black, giddings2006black, preskill1992black, gambini2004realistic}. According to the information loss paradox, the principles of unitarity and quantum physics are violated when quantum data contained within matter that is absorbed by a BH becomes irreversibly lost. The introduction of entropy corrections for BHs, dependent on the Planck length as the basic scale of quantum gravity, offers a potential resolution to this dilemma \cite{garay1995quantum}. For a comprehensive note on the information loss paradox, the soft particles and Faddeev-Jackiw quantization via conformal diagrams, see \cite{dengiz2018note}. Moreover, the recent paper of Pourhassan et al \cite{pourhassan2023information} explores the detection of quantum gravitational corrections in black holes through information theory, particularly by analyzing the Kullback-Leibler divergence. It finds that while these corrections initially increase the divergence with decreasing black hole mass, they eventually decrease beyond a critical mass due to quantum fluctuations dominating the system, shedding light on the scale and dimension dependence of detecting such corrections.

This thesis mainly focuses on rotating  linear dilaton black holes (RLDBHs) which are known for their  potential connection to dark matter, and therefore, act as a possible candidate for addressing fundamental open questions in both cosmology and astrophysics \cite{clement2003linear,tokgoz2017stationary,sakalli2016analytical,berti2015testing,nashed2023slow,damour1990dark,guzman1998scalar,einasto2009dark}. Dark matter \cite{oks2023review} is a hypothetical form of matter believed to account for a substantial portion of the material in the universe, remaining invisible as it does not interact with light. In contrast, the dilaton, a scalar field originating from string theory, is believed to be pervasive in space and may influence the formation of BHs \cite{ketova2021formation}. Recent studies suggest a potential link between the dilaton field and dark matter, offering RLDBHs as a plausible explanation for certain dark matter characteristics. Interactions between the dilaton field and dark matter particles affecting the motion of stars and galaxies, as described in references \cite{de2023galactic, guzman1998scalar}, may explain phenomena such as galaxies rotating at velocities higher than what visible matter can account for. Throughout this work, unless explicitly stated, we employ natural units where G = $\hbar$ = c = $k_{B}$ = 1 and adopt the metric signature ($+,-,-,-$).

The structure of the current thesis is as follows: In Sections \ref{sec:generalinstructions} and \ref{sec:chaptersnadsections}, we present an overview of Hawking radiation and introduce the physical characteristics of non-asymptotically flat (NAF) RLDBH spacetime metrics. In Chapter \ref{sec:paragraphs}, we apply the Parikh-Wilczek quantum tunneling technique \cite{parikh2000hawking}. The subsequent section provides a concise introduction to GUP. In Section \ref{sec:theorems}, we delve into the entropy of RLDBHs and the temperature adjustments brought about by GUP. To conclude, we summarise our research and offer some closing remarks in Chapter \ref{sec:tablesandfigures}.

\newpage
\section{Hawking Radiation}\label{sec:generalinstructions}

In 1974, Stephen Hawking introduced the theoretical concept known as "Hawking radiation" whose foundation lies in the idea that nothing in the universe truly lacks activity \cite{hawking1976black}. Although this concept might initially appear perplexing quantum fields exist in empty space, even in the absence of mass, particles, or quanta of energy. The conventional explanation posits that these fields can generate pairs of "virtual particles," typically comprising a particle-antiparticle pair, which rapidly annihilate one another since they do not need to possess zero energy. However, in the vicinity of a BH, it is conceivable for one of these particles to venture into the BH, becoming lost forever, while the other escapes as Hawking radiation, as described. Nevertheless, it is worth noting that this explanation, although frequently employed, is not all-encompassing. In fact, the description of how gravity affects spacetime as per General Relativity is what underpins the phenomenon of Hawking radiation. According to HUP, it is not possible to know with absolute certainty how much energy a quantum field has or when a given energy can be exactly assigned to it. This is due to the fact that quantum fields in empty space adhere to this principle. The differing gravitational curvatures of spacetime regions cannot reach a consensus on the energy of the quantum fields because a gravitational field bends spacetime and influences the local passage of time. It is this variation in vacuum energy at various points within a BHs gravitational field that leads to the generation of the so-called "virtual particles."

We understand that there is $\frac{1}{2}\hbar \omega$ energy in every wave mode, even in its vacuum state, including extremely high frequencies. Over time, these variations transform into real exit modes. These modes are highly redshifted, and as time progresses, their redshift intensifies. However, even when the redshift reaches an exceptionally high value, there is always a remarkably high-frequency component in the ongoing wave whose vacuum fluctuations shift to the desired frequency. The name "Hawking radiation" is attributed to these redshifted waves that transforms from being virtual to real.

When the Unruh effect and the equivalence principle are applied to the event horizons of BHs, Hawking radiation ensues. An observer approaching a BH's event horizon must accelerate to prevent being drawn in. This accelerating observer perceives a continuous stream of particles emanating from the local acceleration horizon, altering direction, and eventually descending freely back into the BH. This process maintains local thermal equilibrium, signifying that a finite temperature is attained when extending this thermal bath to infinity. This suggests that some of the particles that the horizon emits become outward Hawking radiation instead of being reabsorbed \cite{dewitt1979quantum}.

\section{Physical Traits of RLDBH}\label{sec:chaptersnadsections}
In this section, we provide a concise overview of RLDBH spacetime, which was originally introduced by Clement et al \cite{clement2003linear}.The theory of Einstein-Maxwell-Dilaton-Axion (EMDA) gravity can be thought of as a simplified version of the bosonic part of $D = 4, \; N = 4$ supergravity \cite{wei2013observing}. The action for the EMDA gravity theory reads \cite{clement2003linear}:

\begin{equation}
 S=\frac{1}{16\pi}\int d^4 x\sqrt{-g}\left[
 -\mathcal{R}+2\partial_\mu\phi\partial^\mu\phi
 +\frac{1}{2}e^{4\phi}\partial_\mu{A}\partial^\mu{A}
 -e^{-2\phi}F_{\mu\nu}F^{\mu\nu}
 -\mathbb{A} F_{\mu\nu}\tilde{F}^{\mu\nu}\right], \label{izm1}
\end{equation}
where the pseudoscalar axion field is denoted as $A$, and the dilaton field is represented by $\phi$. The electromagnetic field strengths of $A$ is Abelian vector field this vector field in which the order of vector operations does not affect the result and its dual counterparts are symbolized as $F$ and $\tilde F$, respectively. Additionally, $\mathcal{R}$ corresponds to the Ricci scalar. In addition to the static black hole solution outlined in the EMDA theory \eqref{izm1}, ref. \cite{clement2003linear} provides an explicit metric for RLDBH spacetime that eliminates the need for the NUT charge, as discussed in \cite{hawking1999nut}:

 \begin{eqnarray} \label{izm2}
 ds^2=\frac{\Delta}{r_0r}dt^2-r_0r \left[\frac{dr^2}{\Delta}
 +d\theta^2+\textrm{sin}^2\theta\big(d\phi-\frac{a}{r_0r}dt\big)^2 \right]\, 
 \end{eqnarray}
in which

\begin{equation} \label{izmx2}
    \Delta = r^{2}-2Mr+a^{2},
\end{equation}
and the fields in the background are supplied by

 \begin{eqnarray} \label{izm3}
 &&F=\frac{1}{\sqrt{2}}\left[\frac{r^2-a^2\textrm{cos}^2\theta}{r_0r^2}dr\wedge
 dt+a\textrm{sin}2\theta d\theta\wedge\big(d\phi-\frac{a}{r_0r}dt\big)
 \right], \\
 &&e^{-2\phi}=r_0r \chi,\\
 && {A}=-r_0a\textrm{cos}\theta \chi, 
 \end{eqnarray}
where 
\begin{equation}
    \chi= (r^{2}+a^{2} cos^{2}(\theta))^{-1}
\end{equation}
In the meantime, the spacetime's rotation and background charge are indicated by the physical parameters $a$ and $r_{0}$, respectively. Metric \eqref{izm2} was actually obtained by applying a specific solution-generating method to the Kerr metric. The spacetime metric is essentially the static linear dilaton metric, but it exhibits different behavior as it approaches infinity, as demonstrated by \eqref{izm2}. Its behavior is similarly distinct from the Kerr metric's behavior near $r = 0$. Concerning the Kerr metric, it is possible to extend the metric to negative $r$ by passing through a disc located at $r = 0$. On the other hand, $r = 0$ in Eq. \eqref{izmx2} represents a timelike line singularity. Due to this, the Penrose diagrams of the metric \eqref{izm1} differ from those of the Kerr spacetime but remain the same for all three scenarios ($a^{2} < M^{2}$, $a^{2} = M^{2}$, and $a^{2} > M^{2}$). Instead, they show similarities to the Penrose diagrams of the Reissner-Nordstr\"{o}m care such that charge is replaced by a.

Instead, they bear a resemblance to the Penrose diagrams of the Reissner-Nordstr\"{o}m spacetime, in which the rotation parameter $a$ (or angular momentum) takes the place of the charge.

We shall now describe the thermodynamic characteristics of the RLDBH. It should be recognized right away that $M$, which was present in the solution, is no longer the ADM mass. The mass computation of Brown and York \cite{brown1993quasilocal}, who established the mass for NAF spacetimes, may be used to compute the pertinent thermodynamic quantities in order to derive the first law of BH physics. The following is the relationship between the mass parameter $M$ and the quasilocal mass $\tilde{M}$ of the RLDBH:

\begin{equation}
   M= 2\widetilde{M}, 
\end{equation}
and the angular momentum $J$ is given by
 
 \begin{eqnarray}
 2J= ar_{0}.
 \end{eqnarray}
The statistical Hawking temperature $T_{RLDBH}$, the Bekenstein-Hawking
 entropy $S_{RLDBH}$, and the angular velocity
 $\Omega_{RLDBH}$ of the RLDBH are  given by \cite{sakalli2015quantization}

 \begin{eqnarray}
 &&T_{RLDBH}=\frac{\kappa}{2\pi}=\frac{r_+-r_-}{4\pi r_0 r_+},\label{triplet}\\ 
 &&\Omega_{RLDBH}=\frac{a}{r_0 r_+}, \\
 &&S_{RLDBH}=\pi r_0 r_+, \label{entropy}
 \end{eqnarray}
in which $r_\pm=M\pm\sqrt{M^2-a^2}$ correspond to the positions of the inner and outer (event) horizons, respectively. The first law of BH mechanics can be verified to hold for the RLDBH using the thermodynamical constants described above:

 \begin{eqnarray} d\widetilde{M}=T_{RLDBH}dS_{RLDBH}+\Omega_{RLDBH} dJ \label{iz7}
 \end{eqnarray}
It should be noted that
 the electric charge takes the form
 
 \begin{eqnarray}
 Q=\frac{r_0}{\sqrt{2}}
  \end{eqnarray}
It is worthy to recall that in, Eq. \eqref{iz7}, the quantity $Q$ does not appear. This is characteristic of linear dilaton backgrounds, where the differentiations, as indicated in Eq. \eqref{iz7}, are performed with the electric charge $Q$ held constant. In this context, the background charge is equivalent to the electric charge $Q$. Furthermore, the entropy \cite{carlip2000logarithmic} at extremality is expressed as follows, and the extremality condition is given by $M = a$:
  
  \begin{equation} \label{izex}
 S_{RLDBH}(T=0)=\pi r_{0} a.
\end{equation}

As evident from the previous explanation, extremal BHs have the smallest event horizon area, which in turn means that the entropy linked to these BHs is also minimised. Consequently, the entropy of an extremal BH is usually significantly less than that of a non-extremal BH \eqref{izex}.

\section{Hawking
Radiation of RLDBH via Quantum Tunnelling Technique } \label{sec:paragraphs}
As is well-known, the WKB approximation furnishes information regarding the following tunneling rate at which an $s$-wave tunnels from inside to outside the event horizon of a BH:

\begin{equation}
\Gamma = \Gamma_{0} e^{-2 Im\mathcal{I}}
\label{1}
\end{equation}
where $\mathcal{I}$ represents the action of the tunneling particle, and $\Gamma _{0}$ stands for a normalization factor. Nevertheless, as the radiation emitted from a BH adheres to the Boltzmann distribution in a traditional sense, it is possible to assert the following regarding the rate at which energy particles are emitted from the BH's horizon:

\begin{equation}
\Gamma =\Gamma_{0}\exp(-\beta E),\label{2}
\end{equation}%
where
\begin{equation}
\beta =\frac{1}{T},  \label{3}
\end{equation}
corresponds to the inverse temperature ($T$) and it is known as the Boltzmann constant \cite{hawking1975particle}. Then, the imaginary component of the action for a tunneling particulate can be estimated in terms of a $s$-wave, as mentioned in the Ref. \cite{heidari2023quantum}:

\begin{equation}
Im\mathcal{I}=Im\int_{0}^{p_{r}}
\int_{r_{h}(M)}^{r_{h}(M-E)}dp_{r}dr,  \label{4}
\end{equation}
where $r$ is the distance from the BH's center and $M$ represents the BH's initial mass overall. $M - E$ represents the BH's final mass following the radiation released by the particle that has its energy tunnelled out. The following outcome is obtained by applying Hamilton's equations of motion:

\begin{equation}
\overset{\cdot }{r}=\frac{dH}{dp_{r}}=\frac{d(M-\omega )}{dp_{r}},  \label{5}
\end{equation}%
where

\begin{equation}
dp_{r}=\frac{d(M-\omega )}{\overset{\cdot }{r}}  \label{6}
\end{equation}
If we substitute the above expressions in integral \eqref{4}, we get%

\begin{equation}
Im\mathcal{I}=Im\int_{0}^{E}\int_{r_{h}(M-E)}^{r_{h}(M)}\frac{dr}{%
\overset{\cdot }{r}}d\omega.  \label{7}
\end{equation}

It is worth noting that the majority of the radiation spectrum is often dominated by zero-mass particles since BHs typically have very low Hawking temperature values  \cite{wald2010general}. A tunneling particle with negligible mass travels along a radial path characterized by a null geodesic in the context of an s-wave. The generic metric expression of RLDBH spacetime \eqref{izm2} can be redefined as follows:

\begin{eqnarray}
 ds^{2}&=\frac{r^{2}-2Mr+a^{2}\cos ^{2}\theta }{r_{0}r}dt^{2}-\frac{r_{0}r}{%
r^{2}-2Mr+a^{2}}dr^{2}-r_{0}rd\theta ^{2}\\ \notag
&-r_{0}r\sin ^{2}\theta d\phi
^{2}+2a\sin ^{2}\theta dtd\phi  \label{8}   
\end{eqnarray}
It is reasonable to assume that for an observer at spatial infinity, the radiation from a spinning BH maintains spherical symmetry. This enables us to describe the tunneling process of a spinning BH in terms of the $s$-wave approximation. However, as a particle tunnels through the rotating event horizon, it will interact with and be influenced by the BH's spin. In such circumstances, the tunneled particle will exhibit motion in the $\phi$ direction at a non-zero rate of change, $d\phi \neq 0$. To account for this motion, we can employ a reference frame that rotates in synchronization with the BH's horizon's motion over time. To achieve this, we apply a coordinate transformation based on the rotation, as follows:%

\begin{equation}
\phi =\phi ^{\prime }+\Omega _{h}t,   
\label{9}
\end{equation}%
which leads to

\begin{equation}
d\phi =d\phi ^{\prime }+\Omega _{h}dt,
\end{equation}%
and 

\begin{equation}
d\phi ^{^{2}}=d\phi ^{\prime 2}+2\Omega _{h}d\phi dt+\Omega _{h}^{2}dt^{2},
\end{equation}
where the angular velocity of a spinning BH's event horizon is $\Omega _{h}$, a constant:%

\begin{equation}
\Omega _{h}=\frac{g_{t\phi }}{g_{\phi \phi }}\bigg\rvert_{r=r_{h}}=\frac{a}{r_{0}r_{h}}.
\end{equation}

From this point onward, we will employ the equisymbol "$r_{h}=r_{+}$" to represent the event horizon of the RLDBH. When observers are positioned at this horizon within a rotating reference frame, they will observe that the BH's angular speed, denoted as "$\Omega_{h}$," equals zero: $\Omega_{h}(r{h}) = 0$. This phenomenon arises due to their close proximity to the event horizon, hindering their detection of the black hole's rotation.

Within this co-rotating reference frame, a particle will not experience the gravitational attraction resulting from the BH's spin. This is due to the spontaneous occurrence of particle tunneling at the horizon. Consequently, the particle undergoing tunneling displays no motion in the degrees of freedom represented by $\phi ^{\prime}$. Hence, it is a valid assumption to consider that d$\phi ^{\prime} = 0$, indicating that there is no change in the $\phi ^{\prime}$ coordinate as the particle tunnels through the horizon.

In contrast, when analyzing Hawking radiation using a quantum tunneling approach, we reevaluate the RLDBH metric \eqref{7} within the co-rotating reference system by setting $\theta=0$, representing the equatorial plane. As a result, metric \eqref{7} transforms in the following manner:

\begin{equation}
ds^{2}\mid _{\theta =0}=\frac{r^{2}-2Mr+a^{2}}{r_{0}r}dt^{2}-\frac{r_{0}r}{r^{2}-2Mr+a^{2}}%
dr^{2}.
\end{equation}

On the horizon, $g_{rr}=\frac{r_{0}r}{r^{2}-2Mr+a^{2}}$ is singular, and we
have to remove that coordinate singularity. To accomplish this, we
pass to the Painlev\'{e} coordinate system:%

\begin{equation}
dt=dT-\sqrt{\frac{g_{rr}(r)-1}{G_{tt}(r)}}dr,
\end{equation}%

\begin{equation}
dt^{2}=dT+\frac{g_{rr}-1}{G_{tt}(r)}dr^{2}-2\sqrt{\frac{g_{rr}-1}{G_{tt}(r)}}%
dtdr,
\end{equation}
where \

\begin{equation}
G_{tt}(r)=\frac{(r-r_{-})(r-r_{+})}{r_{0}r},
\end{equation}
which transforms the metric into%

\begin{equation}
ds^{2}=G_{tt}dT^{2}-2\sqrt{G_{tt}(g_{rr}-1)}dTdr-dr^{2}.
\end{equation}
As is well-known, in the case of null geodesics, i.e., $ds^{2}= 0$, we
obtain%

\begin{equation}
\frac{\dot{r}}{X}=\sqrt{G_{tt}(r)g_{rr}(r)}, \label{hp1}
\end{equation}
in which%

\begin{equation}
1-X=\sqrt{1-g^{rr}(r)} \label{hp22}
\end{equation}
After substituting Eqs. \eqref{hp1} in \eqref{7}, the imaginary part of the
tunneling particle's action becomes%

\begin{equation}
Im\mathcal{I}=Im\int_{0}^{E}\int_{r_{h}(M-E)}^{r_{h}(M)}\frac{dr}{X\sqrt{%
G_{tt}(r)g_{rr}(r)}}d\omega
\end{equation}
After multiplying the numerator and denominator parts, separately, by $\eth=2-X$, we obtain

\begin{equation}
Im\mathcal{I}\approxeq Im\int_{0}^{E}\int_{r_{h}(M-E)}^{r_{h}(M)} \sqrt{\frac{G^{tt}(r)}{g_{rr}(r)}} \eth d\omega dr.
\end{equation}

Since $g_{rr}(r)$ exhibits singularity at the horizon, one can express $g_{rr}(r)$ for the metric of a four-dimensional spinning BH as follows:%

\begin{equation}
g_{rr}(r)=Y(r)(r-r_{h})^{-1}, \label{hp2}
\end{equation}
where

\begin{equation}
Y(r)=rr_{0}(r-r_{-})^{-1}. \label{hp3}
\end{equation}

Equation \eqref{hp3} is a function that is regular on the horizon. By substituting Eq. \eqref{hp2} into
Eq. \eqref{hp1}, we find out%

\begin{equation}
Im\mathcal{I}=Im\int_{0}^{E}\int_{r_{h}(M-E)}^{r_{h}(M)}\frac{Y(r_{h})(1+%
\sqrt{1-\frac{r-r_{h}}{Y(r_{h})}})}{r-r_{h}}\sqrt{\frac{g^{rr}(r_{h})}{%
G_{tt}(r_{h})}}d\omega dr. \label{hp4}
\end{equation}

The integral equation \eqref{hp4} now exhibits a pole at $r_{h}$. By adding an extremely small imaginary component to the variable $r$ and enabling the integration path to create a semicircular contour around the pole, it is possible to compute the integration of the radial component. This strategy leads to the following results:

\begin{equation}
\frac{Im\mathcal{I}}{2\pi }=\int_{0}^{E}Y(r_{h})\sqrt{\frac{g^{rr}(r_{h})}{%
G_{tt}(r_{h})}}d\omega \label{hp500}
\end{equation}
It is a reasonable deduction to make that the energy ($E$) of the particle
involved in tunneling is significantly less than the total mass ($M$) of the black
hole; namely, $E$ is much smaller than $M$. Consequently, when we
consider Eq. \eqref{hp500}, we can simplify the expression within the integral
to a constant value. This leads us to the following result:

\begin{equation}
\frac{Im\mathcal{I}}{2\pi E}=Y(r_{h})\sqrt{\frac{g^{rr}(r_{h})}{G_{tt}(r_{h})}}. \label{hp5}
\end{equation}

Since $\left(\mathcal{G}_{t t}(r), g^{r r}(r)\right) \rightarrow 0$ around the event horizon $\left(r_h\right)$, we can expand them as follows%

\begin{eqnarray}
g^{rr}(r) &\approx &g^{rr^{\prime }}(r_{h})(r-r_{h})+..., \label{hp6}\\
G_{tt}(r) &\approx &G_{tt}^{^{\prime }}(r_{h})(r-r_{h})+....\label{hp7}
\end{eqnarray}

where \textquotedblleft\ . . . \textquotedblright\ seen in Eqs. \eqref{hp6} and
\eqref{hp7} present the high-order terms of ($r - r_{h}$) and prime \textquotedblleft
$\prime$\textquotedblright\ symbol denotes the derivative with respect to $r$. From
Eq. \eqref{hp6}, we get%

\begin{equation}
g^{rr^{\prime }}(r_{h})=\frac{1}{Y(r_{h})}. \label{hp8}
\end{equation}

Substituting Eqs  \eqref{hp6} and \eqref{hp7} in \eqref{hp5}, the near-horizon version of the action can be rewritten

\begin{equation}
Im\mathcal{I}\approx \frac{2\pi E}{\sqrt{\frac{G_{tt}^{^{\prime }}(r_{h})}{%
Y(r_{h})}}}. \label{hp9}
\end{equation}

Eventually Eq. \eqref{2}, which stands for computing the tunneling rate \eqref{1} and it is equivalent to Eq. \eqref{3} having the Boltzmann constant, is derived

\begin{equation}
Im\mathcal{I}=\frac{\pi E}{\kappa (r_{h})} \label{hp10}
\end{equation}
By matching Eqs. \eqref{hp9} and \eqref{hp10}, the surface gravity is obtained as

\begin{equation}
\kappa (r_{h})=\frac{\sqrt{G_{tt}^{\prime }(r_{h})Y^{-1}(r_{h})}}{2}, \label{hp11}
\end{equation}
which yields the surface temperature of the RLDBH:%

\begin{equation}
T_{H}=\frac{\sqrt{G_{tt}^{\prime }(r_{h})Y^{-1}(r_{h})}}{4\pi }. \label{hp12}
\end{equation}

Hence, we successfully derived the Hawking temperature \eqref{hp12} utilizing the quantum tunneling method. Additionally, the surface gravity of the RLDBH spacetime can be computed through the implementation of the timelike Killing vector \cite{wald2010general}:%

\begin{equation}
\kappa (r_{h})=\lim\limits_{r\rightarrow r_{h}}\frac{\partial _{r}\sqrt{%
G_{tt}}}{\sqrt{g_{rr}}}=\lim\limits_{r\rightarrow r_{h}}\frac{\partial
_{r}G_{tt}}{\sqrt{4g_{rr}G_{tt}}}. \label{hp13}
\end{equation}

It is widely acknowledged that near the event horizon, $\kappa(r_{h})$ remains constant, as per the principles of BH thermodynamics \cite{hayward1998unified}. Hence, it can be assessed at any angle $\theta_{0}$. By substituting Eqs. \eqref{hp6} and \eqref{hp7} into Eq. \eqref{hp13}, we derive

\begin{equation}
\kappa (r_{h})=\frac{\sqrt{G_{tt}^{\prime }(r_{h})Y^{-1}(r_{h})}}{2},\label{hp14}
\end{equation}
After a concise comparison between Eqs. \eqref{hp11} and \eqref{hp14}, it is evident that both equations yield identical surface gravities. Moreover, it is a firmly established fact that $\kappa (r_{h})$ remains constant at the horizon. Therefore, the specific determination for the surface gravity of the RLDBH, derived from Eq. \eqref{hp14}, should remain unaffected by the parameter $\theta$. Put differently, the surface gravity and the Hawking temperature, expressed in Eqs. \eqref{hp11} and \eqref{hp12}, respectively, will persist unchanged regardless of the value assigned to the parameter $\theta$.

% Begin the SECOND CHAPTER =========================================

\chapter{Generalized Uncertainty Principle } \label{ch:preliminary}
HUP, a fundamental concept in quantum physics, has been historically modified to incorporate minimal-length effects into quantum mechanics \cite{nozari2012minimal,pedram2012new,das2012path}. Consequently, as we approach the Planck scale, the uncertainty in the position operator reaches a global minimum, leading to the emergence of the GUP (see \cite{bosso202330} and references therein for detailed information). GUP extends beyond HUP by integrating concepts from relativistic physics and quantum gravity.

Unlike the conventional uncertainty principle, GUP suggests the potential existence of a fundamental upper limit on accurately determining both position and momentum\footnote{This chapter is primarily cited from \cite{sucu2023gup}}. The notion that our comprehension of space and time might blur or become ambiguous at extremely small scales or high energies is frequently associated with GUP. It proposes a fundamental restriction on our ability to precisely determine particle location and momentum. This concept arises in the effort to reconcile quantum mechanics with general relativity, the theory that explains gravity on a cosmic scale. The primary objective of quantum gravity theories is to provide a unified framework for these two fundamental branches of physics.

In various theoretical frameworks, GUP manifests in different forms. However, a common modification to HUP involves introducing additional terms, besides the reduced Planck constant ($\hbar$), dependent on the Planck length ($l_{P}$) and the speed of light ($c$). A generic form for representing GUP in one-dimensional space is as follows:

\begin{equation}
    \Delta x\Delta p \geqslant \frac{\hbar}{2}(1+\alpha^{2}(\Delta p)^{2}).
\end{equation}

It proves that there is a minimal length $ \Delta x\geqslant \hbar \alpha $, where $\alpha$ is a positive constant independent of the uncertainties associated with momentum and position, $\Delta p$ and $\Delta x $, and $x$ and $p$, respectively. $[x, p]_{GUP}$ = i$\hbar$(1 + $\alpha^{2}$ $p^{2}$) is the commutation relation for GUP, where $p$ and $x$ stand for the momentum operators and positions, respectively. Thus, the modified HUP relation is used to eliminate the divergence in the brick-wall model, as described in \cite{brustein2011black}. Utilising the corrected state density provided by the GUP, the statistical entropy of many BHs has also been computed \cite{zhao2012generalized}. The findings thus demonstrate the finite nature of the statistical entropy of the near-horizon quantum state density.

In contrast, we examine GUP within the framework of tunneling formalism and compute the quantum-corrected Hawking temperature and entropy for a self-dual BH using the Hamilton-Jacobi technique. Now, let us commence our investigation using the GUP \cite{ali2009discreteness,cai2008corrected}:

\begin{equation}
    \Delta x\Delta p \geqslant \hbar (1-\frac{\alpha l_{P}}{\hbar}\Delta p + \frac{\alpha^{2} l_{P}^{2}}{\hbar^{2}}(\Delta p)^{2}),\label{2.2}
\end{equation}
in which $l_{P}$ is the Planck length ($\approx$ $10^{-35}m$). Now, Eq. \eqref{2.2} can be written as follows:

\begin{equation}
    \Delta p _{GUP} \geqslant \frac{\hbar(\Delta x + \alpha l_{P})}{2 \alpha^{2} l_{P}^{2}}  \left (1- \sqrt{1- \frac{4 \alpha^{2} l_{P}^{2}}{(\Delta x + \alpha l_{P})^{2}}} \right ),
\end{equation}
where the negative sign is our choice. Considering that $\frac{l_{P}}{\Delta x}$ is significantly smaller in comparison to unity, we can expand the preceding equation using the Taylor series:

\begin{equation}
\Delta p_{\mathrm{GUP}} \geqslant \frac{1}{2 \Delta x}\left[1-\frac{\alpha}{2 \Delta x}+\frac{\alpha^2}{2(\Delta x)^2}+\ldots\right],
\label{2.4}
\end{equation}

in which $l_{P}$ and $\hbar$ are set to 1, so GUP becomes

\begin{equation}
    \Delta x\Delta p \geqslant 1
\end{equation}
Proceeding now with the saturated version of the uncertainty principle, one gets

\begin{equation}
    \xi \Delta x \geqslant 1
\end{equation}
Thus, given the saturated version of HUP and $\Delta x$ $\Delta p$ $\geqslant $ 1, where $\xi$ is the quantum particle's energy, Eq. \eqref{2.4} becomes

\begin{equation}
    \xi_{QGC} \geqslant \xi \frac{1}{2 \Delta x} \left [ 1- \frac{\alpha }{2 \Delta x} +  \frac{\alpha^{2} }{2 (\Delta x)^{2}} +... \right].
\end{equation}

The quantum tunneling rate for a quantum particle with $\xi_{QGC}$

\begin{equation}
    \Gamma \simeq \exp{(-2Im \mathcal{I}) } = \exp{(\frac{\xi_{QGC}}{T_{QGC}})},\label{2.8}
\end{equation}
where QGC temperature is $T_{QGC}$. Comparing Eq. \eqref{2.8} with the Boltzmann factor now yields

\begin{equation}
    T_{QGC} \geqslant T_{H} \left [ 1- \frac{\alpha }{2 \Delta x} +  \frac{\alpha^{2} }{2 (\Delta x)^{2}} +... \right]^ {-1}.
\end{equation}
Thus, one can obtain QGC-entropy \cite{sakalli2016gup} by applying the law of BH thermodynamics:

\begin{equation}
    S_{QGC}  =\int\frac{\kappa dA_{h}}{8\pi T_{QGC}}=\int\frac{T_{H}dA_{h}}{4T_{QGC}}.
\end{equation}

\section{GUP-corrected Temperature and Entropy of RLDBH }\label{sec:theorems}

HUP is expanded upon by GUP, which considers the implications of quantum gravity \cite{carr2015sub,carr2022generalized}. It implies that there is a minimal length that may be seen, which changes the uncertainty connections between the momentum and position. Researchers have been examining how GUP affects a number of fundamental phenomena, such as the thermodynamics of BH's, in recent years. The role of GUP corrections in the entropy evaluations of charged and revolving BHs have been an active research topic \cite{gecim2018gup,gecim2018quantum}. Therefore, in this thesis, the influence of such corrections on RLDBH entropy will be examined. In light of this, let us examine the Lense-Thirring effect, a gyroscopic precession that is observable \cite{bardeen1975lense} and which may be acquired by the dragging coordinate transformation \eqref{9} that produces the following metric:

\begin{equation}
ds^{2}=\frac{r^{2}-2Mr+a^{2}\cos^{2}\theta}{r_{0}r}dt^{2}-\frac{r^{2}-2Mr+a^{2}}{r_{0}r}dr^{2}-r_{0}r\sin^{2}\theta d\phi'^{2}-r_{0}rd\theta^2 \label{51}%
\end{equation}
For a scalar field, the Klein-Gordon equation (KGE) with GUP takes the following form: \cite{sakalli2022topical,todorinov2019relativistic,jusufi2023generalized}.

\begin{equation}
-(i\hbar)^{2}\partial^{t}\partial_{t}\Psi=[(i\hbar)^{2}\partial
^{i}\partial_{i}+m_{p}^{2}]\times\lbrack1-2\alpha_{GUP}(i\hbar)^{2}%
\partial^{i}\partial_{i}+m_{p}^{2}]\Psi 
 \label{54}%
\end{equation}
 In this context, $m_{p}$ represents the mass of the scalar particle, while $\alpha_{GUP}$ refers to the GUP parameter. The semi-classical WKB approximation technique can be utilized to address the generalized KGE \eqref{54} as described in reference \cite{ovgun2016massive}. For this purpose, one can employ the following assumption:

\begin{equation}
\Psi(t,r,\theta,\phi)=\exp\left(\frac{i}{\hbar}S(t,r,\theta,\phi)\right), \label{55}%
\end{equation}
in which $S(t,r,\theta,\phi$) represents the forbidden tunnelling action. For the action, we can utilize the following Hamilton-Jacobi ansatz \cite{mirekhtiary2019hawking,sucu2023gup} to account for the symmetries of metric \eqref{51}:

\begin{equation}
S(t,r,\theta,\psi)=-Et+W(r)+K(\theta)+j\phi+C, \label{56}%
\end{equation}
where $E$ represents energy, $j$ represents the particle's angular momentum, and $C$ is a complex constant. In the leading order of $\hbar$, the following is obtained by substituting action \eqref{56} in Eq. \eqref{55}:

\begin{multline}
    \frac{r_{0}r}{r^{2}-2Mr+a^{2}}E^{2}=\frac{r^{2}-2Mr+a^{2}}{r_{0}r}W^{\prime
}(r)^{2}+\frac{j^{2}}{r_{0}r\sin^{2}\theta}+\frac{{K^{\prime}(\theta)}^2}{r_{0}r}\\
+m_{p}^{2}\left[1-2\alpha
_{GUP}\left(\frac{r^{2}-2Mr+a^{2}}{r_{0}r}\right)W^{\prime}(r)^{2}-\frac{2\alpha_{GUP}j^{2}%
}{r_{0}r\sin^{2}\theta}-   \frac{2\alpha_{GUP}K^{\prime}(\theta)^{2}%
}{r_{0}r}   -2\alpha_{GUP}m_{p}^{2}\right]. \label{57}%
\end{multline}

Therefore, the integral solution obtained from the radial part (which ignores the higher-order elements of ($\alpha_{GUP}$) is as follows:

\begin{equation}
W(r)=\pm\int\frac{Edr}{\sqrt{\Delta(1-2\alpha_{GUP}m_{p}^{2})}}, \label{58}%
\end{equation}
Remember that $\Delta$ was defined as $\Delta= r^{2}-2Mr+a^{2}$. Subsequently, the path of the contour can be modified to compute the integral around the singularity at $r_{h}$, which allows us to obtain the following result:

\begin{equation}
W(r_{+})=\frac{i\pi E}{2\sqrt{1-2\alpha_{GUP}m_{p}^{2}}}\frac{r_{0}r_{+}}%
{\sqrt{M^2-a^2}}. \label{61}%
\end{equation}

Consequently, the following method can be used to calculate the RLDBH's Hawking temperature using GUP:

\begin{equation}
T^{GUP}_{RLDBH}=\frac{\sqrt{M^2-a^2}}{2\pi r_{0}r_{+}}\sqrt{1-2m_{p}^{2}\alpha_{GUP}}.
\label{62}%
\end{equation}

\begin{figure}[!htbp]
    \centering 
    \includegraphics[scale=0.5]{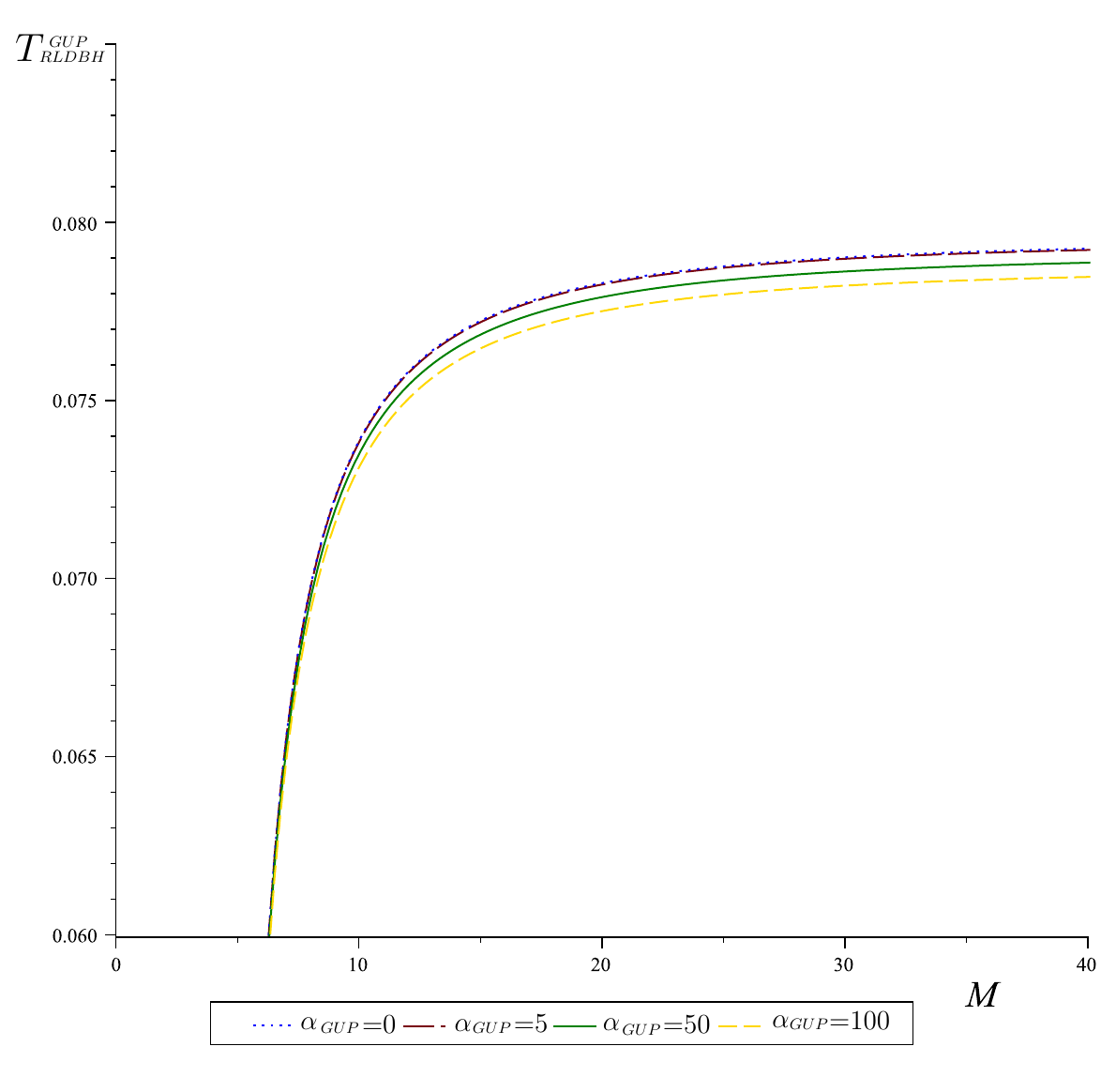}
        \caption{\centering $T^{GUP}_{RLDBH}$ versus $M$ plots for various $\alpha_{GUP}$ values. The plots are determined by Eq. . The physical parameters are chosen as $a=5$, $m_{GUP}=0.01$, and $r_{0}=1$. }
    \label{fig:disc1} \label{FFig1}
\end{figure}

In relation to the preceding formula, as $\alpha_{GUP}$ approaches zero, the GUP-modified Hawking temperature reverts to the original Hawking temperature \eqref{triplet}. Figure \ref{FFig1} illustrates the growth of GUP effect ($\alpha_{GUP}$) concerning BH mass and its efficacy in reducing the Hawking temperature.

Quantum physics textbooks typically derive the standard HUP ($\Delta x\Delta p\geqslant1$) and its fully realized version, as discussed in references \cite{anacleto2015quantum,anacleto2015quantumb}, in scenarios where the GUP effect is absent ($\alpha_{GUP}$ = 0).

\begin{equation}
\xi\Delta x\geqslant1 ,\label{64}%
\end{equation}
The energy of the quantum-scale particle is represented by $\xi$. However, the energy associated with quantum gravity corrected (QGC) is obtained via Refs.\cite{qi2019quantum,gecim2018quantum} by accounting for the GUP:

\begin{equation}
\xi_{QGC}\geqslant\zeta\lbrack1-\frac{\alpha_{GUP}}{2\Delta x}+\frac
{\alpha_{GUP}^{2}}{(\Delta x)^{2}}+...]. \label{65}%
\end{equation}

Referring to Anacleto et al.'s publications \cite{anacleto2015quantum,anacleto2015quantumb}, one can estimate the quantum tunneling rate of a quantum particle with $\xi_{QGC}$ by

\begin{equation}
\Gamma\simeq\exp \left[-\operatorname{Im}S\right]=\exp \left[-\frac{\xi_{QGC}}%
{T_{QGC}}\right], \label{66}%
\end{equation}
in which $T_{QGC}$ represents the QGC temperature and it reads

\begin{equation}
T_{QGC}=T_{H}\left[1-\frac{\alpha_{GUP}}{2\Delta x}+\frac{\alpha_{GUP}^{2}}{(\Delta
x)^{2}}+...\right]^{-1} \label{67}%
\end{equation}
In light of the ongoing investigations \cite{anacleto2015quantum,anacleto2015quantumb}, the alteration in $x$, denoted as $\Delta x$, can be attributed to $\frac{A_{h}}{\pi}$, where $A_{h}$ represents the area of the event horizon. Consequently, applying the first law of BH thermodynamics, the calculation of the GUP corrected entropy is as follows:

\begin{align}
S^{GUP}_{RLDBH}  &  =\int\frac{\kappa dA_{h}}{8\pi T_{QGC}}=\int\frac{T_{H}dA_{h}%
}{4T_{QGC}}\label{68}\\
&  =\int\frac{dA_{h}}{4}[1-\frac{\pi\alpha_{GUP}}{2A_{h}}+\frac{\pi^{2}%
\alpha_{GUP}^{2}}{A_{h}^{2}}+...]\nonumber\\
&  =S_{RLDBH}-\frac{\pi\alpha_{GUP}}{8}\ln(4\pi r_0 r_+)-\frac{\pi^{2}%
\alpha_{GUP}^{2}}{32\pi r_0 r_+}+..., \nonumber
\end{align}
wherein $S_{RLDBH}$ was given in Eq. \eqref{entropy}. We now wish to underscore the significance of this modified entropy. It is widely accepted that a BH's entropy is intricately linked to the count of microstates corresponding to a specific macroscopic configuration. With the Hawking radiation exhibiting apparent thermal characteristics and lacking specific correlations with the information that fell into the BH, the information loss paradox becomes relevant \cite{chen2015black,di2023soft}. This raises the question of whether the radiation truly carries information about the initial condition of the material that formed the BH or if it has been irretrievably lost. This conflict challenges the fundamentals of quantum physics, which dictate the continuous preservation of information.

\begin{figure}[!htbp]
    \centering 
    \includegraphics[scale=0.5]{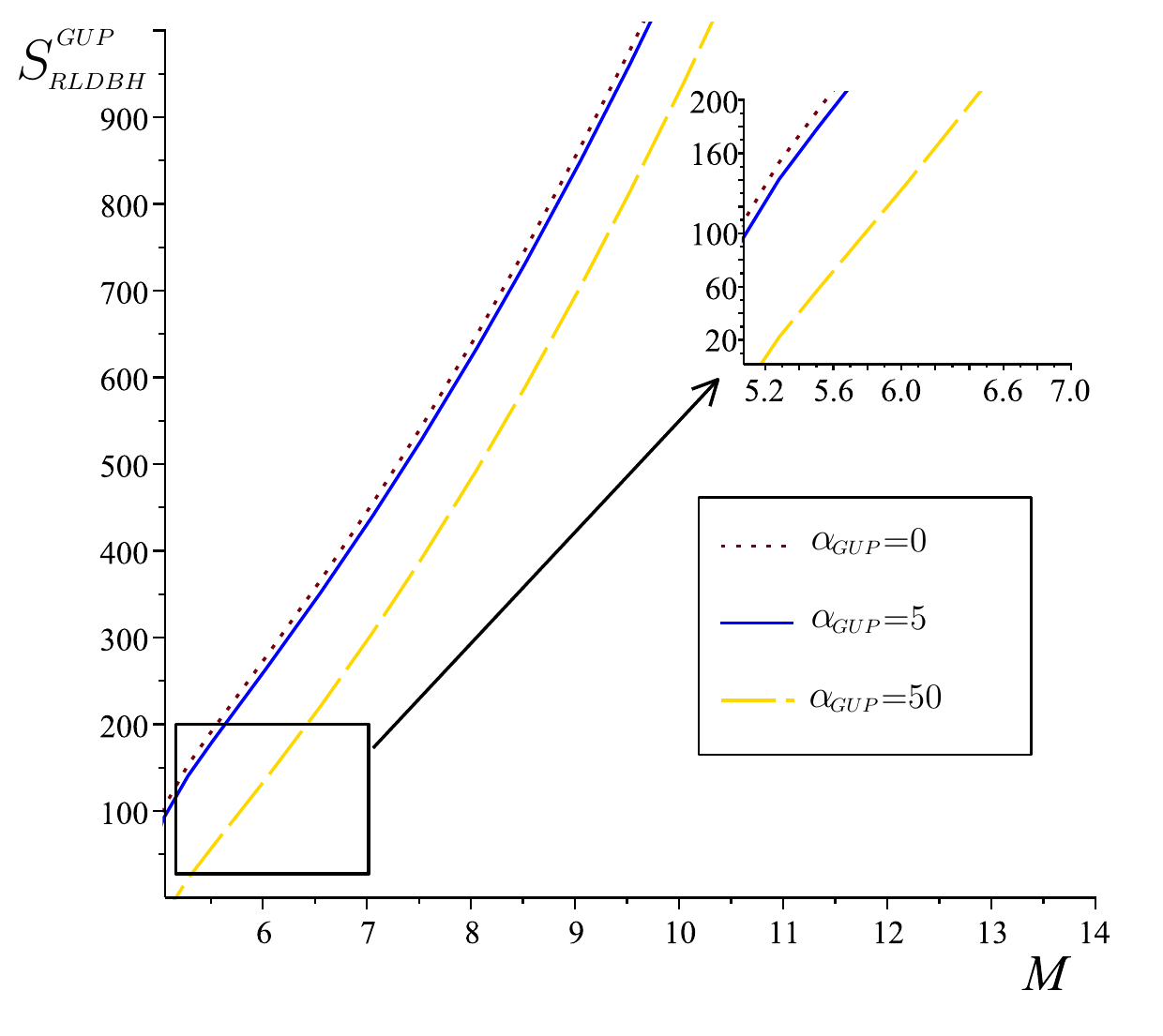}
        \caption{\centering $S^{GUP}_{RLDBH}$ versus $M$ plots for various $\alpha_{GUP}$ values. The plots are determined by Eq. (2.13). The physical parameters are selected as $a=5$, $m_{GUP}=0.01$, and $r_{0}=1$.}
    \label{fig:disc2} \label{fig1}
\end{figure}

The inclusion of GUP in the calculation of BH entropy provides an updated description of the microstates accessible to a BH. This modification offers a potential resolution to the information paradox by suggesting that BHs might retain certain remnants or indications of the absorbed information \cite{chen2015black,sakalli2011entropy}. Figure \ref{fig1} unmistakably demonstrates that, in comparison to a BH with the same mass (RLDBH), entropy decreases as the $\alpha _{GU P}$ parameter increases. However, this reduction in entropy elevates the possibility of information leakage from the BH. This finding concerning RLDBH further corroborates our earlier study \cite{sakalli2011entropy} on the static linear dilaton black hole (SLDBH). The study indicated that quantum gravity effects are crucial not only in resolving the information paradox for the SLDBH but also in establishing a nonzero statistical correlation.

In essence, the physical significance of GUP-corrected entropy lies in its potential to provide insights into the behavior of BHs concerning quantum physics. It provides a way out of the information conundrum and increases our understanding of spacetime and gravity's basic quantum characteristics.

\chapter{Summary and Conclusions}\label{sec:tablesandfigures}
The quantum thermodynamics of RLDBH has been explored within this thesis. The primary objectives of this investigation were twofold: 1) to compute the alterations in temperature and entropy associated with RLDBH due to GUP, and 2) to determine the Hawking temperature of the RLDBH using the null-geodesic tunneling approach elucidated by Parikh and Wilczek \cite{parikh2000hawking}. To achieve those aims, we have employed the technique of dragging coordinate systems. Specifically, we have reduced the four-dimensional spacetime of a spinning BH to a three-dimensional slice within that particular coordinate system \cite{ma2008hawking}. To preserve the original structure of spacetime, we have utilized an event horizon-following coordinate system. This method effectively nullifies the influence of the angular parameter $\phi$ of a tunneling particle. The obtained results indicate that the temperature of thermal radiation emitted the RLDBH, as determined by the quantum tunneling technique, aligns with the statistical Hawking temperature \eqref{triplet}. Consequently, we have successfully demonstrated the complete segregation of the Klein-Gordon equation associated with GUP within the framework of large scalar field propagation in the geometry of RLDBH. This segregation has been achieved using the Hamilton-Jacobi approach. Subsequently, we have revisited the quantum tunneling framework to accurately compute the Hawking temperature for RLDBH, modified by GUP, as expressed in Eq. \eqref{2.8}. Additionally, we have established the Hawking temperature for QGC, as depicted in Eq. \eqref{54}, utilizing the entropy derived from GUP \cite{faizal2015gup}, outlined in Sec. \ref{sec:theorems}. Notably, both temperatures converge to the standard Hawking temperature \eqref{triplet} when the $\alpha_{GUP}=0$ condition is met. Building upon our previous research \cite{sucu2023gup} on the RLDBH, it was illustrated in Fig. \ref{fig1} that $S^{GUP}_{RLDBH}$ decreases with an increasing $\alpha_{GUP}$ parameter. This implies a heightened potential for extracting additional information from the respective BH. 

Overall, our findings demonstrate that Hawking radiation of the RLDBH is quantitatively affected by the GUP, potentially introducing subtle features in the energy spectrum or other radiation properties. Future space-based observatories could help discover GUP effects by enabling more thorough studies of BHs and their radiation, like the Laser Interferometer Space Antenna (LISA) project \cite{amaro2017laser}.

%================== REFERENCES ==================
% You may either use thebibliography environment to manually
% add the references
%\begin{thebibliography}{2} 
%\bibitem{igsr} "Writing Your Thesis, Defence and Graduation Procedures", \textit{Institute of Graduate Studies and Research - EMU}, 2020. [Online]. Available: %https://grad.emu.edu.tr/en/academic-issues/theses. [Accessed: 22- Jun- 2020].

%\bibitem{overleaf} "Overleaf, Online LaTeX Editor", \textit{Overleaf}, 2020. [Online]. Available: https://www.overleaf.com. [Accessed: 22- Jun- 2020].
%\end{thebibliography}

% or you may use a bibtex file with proper style
\bibliographystyle{IEEEtran}
\bibliography{references}

%############ End of Document (don't delete) ##############
\end{document}